# A novel molecule generative model of VAE combined with Transformer for unseen structure generation

**Yasuhiro Yoshikai**[1,*] **Tadahaya Mizuno**[2,*,†]

**Shumpei Nemoto**[1] **Hiroyuki Kusuhara**[1]

[1] Laboratory of Molecular Pharmacokinetics, Graduate School of Pharmaceutical Sciences, The University of Tokyo, 7-3-1 Hongo, Bunkyo, Tokyo, Japan
[2] Laboratory of Molecular Pharmacokinetics, Graduate School of Pharmaceutical Sciences, The University of Tokyo, 7-3-1 Hongo, Bunkyo, Tokyo, Japan, tadahaya@gmail.com
[†]Author to whom correspondence should be addressed.
[*]These authors contributed equally.

**Abstract**

Recently, molecule generation using deep learning has been actively investigated in drug discovery. In this field, Transformer and VAE are widely used as powerful models, but they are rarely used in combination due to structural and performance mismatch of them. This study proposes a model that combines these two models through structural and parameter optimization in handling diverse molecules. The proposed model shows comparable performance to existing models in generating molecules, and showed by far superior performance in generating molecules with unseen structures. Another advantage of this VAE model is that it generates molecules from latent representation, and therefore properties of molecules can be easily predicted or conditioned with it, and indeed, we show that the latent representation of the model successfully predicts molecular properties. Ablation study suggested the advantage of VAE over other generative models like language model in generating novel molecules. It also indicated that the latent representation can be shortened to ~32 dimensional variables without loss of reconstruction, suggesting the possibility of a much smaller molecular descriptor or model than existing ones. This study is expected to provide a virtual chemical library containing a wide variety of compounds for virtual screening and to enable efficient screening.

**Keywords**: molecule generation, Transformer, VAE, latent representation, virtual chemical library

## 1 Introduction

Machine learning has advanced research in various fields of science, and drug discovery is one of them. In low molecular drug discovery, it is said that there can be as many as $10^{60}$ possible drug candidate molecules[1], while only $\sim 10^{9}$ of them are recognized and registered in current databases[2–4]. Accessing this large number of unexplored molecules is expected to be useful for discovering new drug seeds. Therefore, deep learning has been actively used to generate novel molecules in recent years. In particular, the variational autoencoder (VAE)[5] architecture is one of the most popular models since its study by Gomez-Bombarelli et al.[6]. This architecture consists of an encoder and decoder, where the encoder learns to convert the input data into a latent representation, and the decoder learns to restore the original input from the latent representation. In the generation phase, the decoder generates new data from randomly generated latent representations. The advantage of VAE over other generative architecture, such as GAN, language model, or diffusion model is that it generates molecules from the latent representation, and therefore the property of the generated data can be easily predicted from that representation. It also enables us to generate data which have desired properties by learning the posterior distribution of the latent representation conditional on the properties. In molecule generation, it is difficult to use the raw compound structure as the input and the output of the model, and their string representation of them are used in many cases. For example, Gomez-Bombarelli et al.[6] represented the structure of a molecule as a simplified molecular input line entry system (SMILES), and processed them with an encoder and decoder of a recurrent neural network (RNN), a model used in the field of natural language processing (NLP), to learn and represent the distribution of molecules. Elsewhere, Kusner et al.[7] used representation of molecules as a sequence of SMILES generation rules.

In these models, RNN models in NLP are used to encode and decode molecules that are expressed as strings. On the other hand, the Transformer model[8] is now prevalent in NLP as a highly powerful model applicable to various tasks. While the original Transformer model has shown high performance in sentence translation, the Transformer models pretrained with large language corpora, such as BERT[9] or GPT[10], have shown high performance in various tasks such as sentence summarization or conversation response generation. Therefore, it is natural to expect that introduction of the Transformer into VAE as its encoder and decoder will lead to a higher generative performance. In practice, however, the Transformer model is rarely combined with VAE[11] possibly for two reasons. The Transformer model receives and outputs a matrix of variable length called memories, whereas VAE architecture requires the encoder and decoder to transform inputs into a distribution of fixed-dimensional latent variables and restore the input from it, and natural form of Transformer does not fit VAE. Additionally, since Transformer is a powerful model, using such a powerful model as the backbone of VAE is prone to posterior collapse[12], where the decoder ignores the latent variables when generating the output. When posterior collapse occurs, the encoder always outputs the prior distribution of $z$ ($p(z)$), whereas the decoder always outputs the distribution of input data $x$ ($p(x)$) regardless of the latent representation. In this case, the distribution $p(x)$ generated by the decoder is what the observed data follows itself, and the model becomes the language model when it generates variable-length data one by one, such as RNN and Transformer.

In this paper, we propose methods to overcome these two challenges and utilize Transformer as the encoder and decoder of VAE to handle molecules. First, to provide Transformer with a latent variable layer of constant dimension, multiple pooling processes are applied to memory, and in the decoder, the latent representation is added to the embedding of each token during decoding, following previous research in NLP[13]. To prevent posterior collapse when handling diverse molecules, a regularization term to maximize the variational lower bound of the mutual information between the VAE input $x$ and the latent representation $z$ was added to the loss function according to Alemi et al.[14]. The developed VAE model with Transformer as its backbone showed competitive or better performance than existing models, especially in generating novel molecules or scaffolds that do not exist in the learned training set. The effect of regularizing mutual information was evaluated, and it was shown that the model succeeded in balancing the decoding and the regularization of the latent space. Ablation study implied that this high ability to generate novel structure is a characteristic feature of VAE. It was also shown that the dimension of latent variables can be reduced to ~32 without loss of reconstruction accuracy, suggesting that molecules can be represented by much shorter descriptor than existing ones. The latent variables of VAE showed competitive performance in the prediction of various molecular properties. Furthermore, the attention intensity of the encoder was visualized to show the important structures for generating latent representations.

## 2 Related Work

### 2.1 VAE-based molecule generation

VAE is one of the most prevalent models used for generative tasks of deep learning, and has also been applied to de novo drug design. The VAE model is trained with an unlabeled dataset and assumes the probability distribution of the observed data $x$, supposing that the distribution of $x$ is determined by another latent variable $z$. The model consists of an encoder and decoder, and the encoder receives the observed data $x$ and predicts the posterior distribution of $z$, whereas the decoder restores the original data $x$ from latent variables $z$ sampled from the predicted distribution. Most of the studies on molecular VAE represent molecules as strings called SMILES, while some studies on VAE for molecular generation deal with graph representation of molecules as input and output [15,16]. The founder of the SMILES-based strategy was Gómez-Bombarelli et al. [6], who introduced VAE to de novo drug design, using SMILES as the input and the output of VAE and RNN, which can process strings of variable length, as backbones of the encoder and decoder. This strategy has been widely accepted and various applications have been studied [17]. For example, Lim et al.[18] designed a conditional variational autoencoder of molecules, and generated molecules which satisfied conditions of drug likeliness. Chenthamarakshan et al.[19] also achieved the controlled generation of molecules and designed drug candidates for COVID-19. Samanta et al.[20] applied the latent representations of molecules in VAE to measure molecular similarities. Kusner et al.[7] focused on grammatical constraint of SMILES to maintain the high validity of the generated molecules and trained VAE with grammatical parse tree of SMILES.

### 2.2 Molecule generation with Transformer

Transformer has now become one of the most widespread architectures in deep learning. Although this model was originally designed for translation of sentences, it has recently been applied to many tasks related to natural language processing (NLP), and has achieved high performance[9,10]. These prominent results have encouraged scientists to adopt Transformer to SMILES strings to generate molecules for de novo drug design. However, most of these studies were based on language model rather than VAE. A language model is trained to simply predict the probability distribution of the token at each position from the previous sequence, regardless of the latent variables. Bagal et al.[21] developed a language model of SMILES conditioned on molecular properties based on the Transformer decoder, and succeeded in generating molecules which satisfy multiple molecular properties or scaffold conditions. Hong et al.[22] designed a combined model of Transformer and feed-forward network and trained it as a language model. The model not only showed competitive performance on molecular generation, but also generated some drug candidates for COVID-19. Wang et al.[23] distilled a Transformer-based language model into RNN and trained the distilled model to generate molecules on multiple constraints using reinforcement learning. Cofala et al.[24] invented an autoregressive method to generate molecular graphs directly using the attention structure of Transformer.

### 2.3 Combined approach of VAE and Transformer

Despite the popularity of VAE and Transformer in the field of de novo drug design, Transformer-based VAE have not been frequently studied[11]. One of the possible reasons is that while VAE requires a latent bottleneck layer of a fixed dimension, the decoder of Transformer cannot output a sentence from the small fixed-dimensional latent vector of VAE because it requires memory, the encoded information of the input of variable length from the encoder of the input. Kim et al.[25] sampled multiple latent vectors from the posterior distribution estimated by the encoder and used them as the memory of Transformer. Dollar et al.[11] designed a Transformer VAE model whose latent variables from the encoder were low-dimensional memory of fixed length, and the model deconvoluted it to the model size at the decoding part. In the field of NLP, Fang et al.[13] compared several methods to decode a sentence from a latent vector of fixed length using Transformer, and demonstrated that adding the vector to the embedded input of the decoder showed the best performance. Our model is based on this best architecture for chemical language models.

## 3 Method

## 3.1 Model architecture

**Figure 1** illustrates the structure of the model. The model was fed with randomized SMILES of molecules and trained to decode canonical SMILES of them. To make the latent representation contain more essential structural information, different representations of molecules (*randomized* and *canonical* SMILES) were used at the input and the output as in neural machine translation, inspired by Winter R. et al[26]. The embedded randomized SMILES was added with sinusoidal positional encoding and then inputted into the Transformer Encoder. We followed the vanilla Transformer[8] when setting hyperparameters (**Supplementary Table 1**) with some exceptions: there were 8 attention and feed-forward layers in the encoder, and the pre-LN structure[27] was adopted, which means layer normalization was placed before the attention / feed-forward layer and residual connection. The mean and maximum of the *memory* (the output of the encoder) were pooled and concatenated with the memory corresponding to the initial token. The mean and variance of the posterior distribution of latent variables were estimated from the pooled memory. The reparametrized latent variables were added to the embedded input of the decoder. The decoder has the same structure and hyperparameters as the encoder. We used teacher forcing [28] during the training, and calculated loss function as follows:

$$\boldsymbol{\mathcal{L}} = \mathbb{E}_{q_{\phi}(z|x)}[\log p_{\theta}(x|z)] + \beta D_{KL}\left(q_{\phi}(z|x)||p_{\theta}(z)\right)$$

where $\boldsymbol{x}$ is the sampled SMILES, $\boldsymbol{z}$ is the latent variable, $\boldsymbol{\theta}$ and $\boldsymbol{\phi}$ are the parameters to model the encoder($\boldsymbol{q}$) and decoder($\boldsymbol{p}$). $\boldsymbol{\beta}$ is $\mathbf{1}$ in usual VAE, but we set $\boldsymbol{\beta} = \mathbf{0.01}$ to prevent posterior collapse. Alemi et al.[14] proved that reducing the coefficient of the $\boldsymbol{D_{KL}}$ is equal to adding normalization to maximize the variational lower bound of mutual information between the input and the latent variables to the loss function.

## 3.2 Training of the Transformer VAE

### Dataset

We used two different datasets to train the Transformer VAE: Molecular Sets (MOSES)[29] and ZINC-15[2]. MOSES dataset is prepared to evaluate the ability to generate drug-like molecules, and contains approximately 1.9M molecules extracted from ZINC clean leads datasets [2], and its molecules satisfy multiple conditions as a drug candidate, such as having proper molecular weight or logP. MOSES have 1.6M molecules in the training set, 176k molecules in the test set and 176k molecules in the testSF set. Molecules in the testSF set have scaffolds which do not appear in the training or test set and are used to measure the ability of the model to generate novel scaffolds. *Canonical* and *randomized* SMILES of molecules in the training set were generated as the input and target of the model. Here, one molecule can be translated into multiple SMILES, and *canonical* SMILES was chosen from them with a fixed protocol, whereas *randomized* SMILES was randomly chosen from possible SMILES by renumbering atoms in molecules[30,31]. These SMILES were tokenized with the vocabulary shown in **Supplementary Table 2**, which consists of element symbols and characters representing bonds. To iterate the evaluation frequently, 10,000 molecules were randomly sampled from the MOSES test set, and used to evaluate the reconstruction ability of the model.

ZINC-15 is a larger dataset which contains about 1 billion commercially available molecules. This dataset was used to train the VAE to learn and generate broader range of molecules. We sampled approximately 30M molecules from ZINC-15 in a stratified manner with respect to the length of their SMILES[32]. Molecules with other than organic atoms (H, B, C, N, O, F, P, S, Cl, Br and I) or those with SMILES longer than 120 letters were removed. Random/canonical SMILES were generated and tokenized in the same way as MOSES dataset. About 9,000 molecules were sampled as the test set, and the remaining molecules were used for training. All of these processes were conducted using RDKit module version 2023.03.1[33].

### Learning procedure

We set batch size to 128, and one epoch of training set of MOSES and ZINC-15 amounted to about 12k and 240k steps, respectively. Warm up scheduler was adopted as in the study of vanilla Transformer[8]. The detailed hyperparameters are listed in **Supplementary Table 1**. The model was trained for 250k steps.

## 3.3 Evaluation

### Performance as a generative model

30,000 latent vectors were randomly generated following prior distribution $\mathcal{N}(0, I)$, and SMILES were generated from them with the decoder by 4 beam search. For the molecules generated from the model trained with MOSES datasets, we measured the metrics provided by the dataset [29], and compared them with the previously reported scores of other models. *Valid* is ratio of valid SMILES in all generated SMILES. *unique@1000* is ratio of SMILES which appear only once in 1000 valid generated SMILES. *unique@10000* is the same uniqueness in 10000 valid generated SMILES. *FCD* (Fréchet ChemNet Distance) is difference of generated molecules to test set in terms of distribution of molecules in the last layer activations of ChemNet. *SNN* (nearest neighbor similarity) is average similarity of each generated molecule to its nearest molecule in test set. *Frag* (fragment similarity) is cosine similarity of frequency of fragments between generated molecules and test set. *Scaf* (scaffold similarity) is cosine similarity of frequency of scaffolds between generated molecules and test set. *FCD*, *SNN*, *Frag* and *Scaf* metrics were calculated for both test and testSF set. *IntDiv* (internal diversity) is average Tanimoto distance of ECFP (R=2) of generated molecules. *Filters* is ratio of valid unique molecules which satisfied conditions used to filter training & test set. *Novelty* is ratio of valid unique molecules which does not appear in training set. As for the model trained with ZINC-15, only *Valid*, *unique@1000*, *unique@10000*, *IntDiv* and *Novelty* against the training set of the generated molecules were measured because the other metrics[29] depend on the training and test set. Training and generation of molecules were conducted 3 times in the baseline condition and once in the ablation studies.

### Reconstruction and regularization

The reconstruction ability and normalization of the latent space during were measured. Randomized SMILES of test set molecules were encoded and decoded to their canonical SMILES, and their complete reconstruction accuracy (perfect accuracy) and character-wise reconstruction accuracy (partial accuracy)[34] were calculated. Greedy decoding was used for reconstruction.

The distribution of estimated mean and variance of the posterior distribution of the latent variables were calculated at step 10000, 50000, 100000, 250000. Our model yields mean ($\boldsymbol{\mu}$) and log variance ($\mathbf{log}(\boldsymbol{\sigma}^2)$) of the latent variables, and the distributions of them for all elements and all molecules in the test set were calculated.

### Molecular property prediction from latent variables

We used the property datasets of molecules in MoleculeNet [35] provided by the DeepChem module [36] (summarized in **Supplementary Table 3**). We filtered the molecules in the dataset in the same way as the training set. Additionally, we removed the salts of the molecules because the molecules in the training set (both MOSES and ZINC-15) do not contain salts. We substituted the mean of all predictions to the prediction on excluded molecules. The recommended splitting algorithms in MoleculeNet and DeepChem were used, and XGBoost[37] was used to predict the properties of molecules from the means of latent variables estimated from the encoder fed with randomized SMILES. The hyperparameters of XGBoost were optimized by Bayesian optimization with optuna[38] by 100 trials in the ranges shown in **Supplementary Table 4**. ECFP[39] and CDDD[26] were used as the baseline descriptors, and Uni-Mol[40] was used as the baseline model. Note that Uni-Mol is a fine-tuning model, whereas others, including our model, are descriptor generation methods.

## 3.3 Visualization of attention

We computed and visualized the allocation of attention to each molecule in the model trained with MOSES dataset. At each layer of the encoder, attention to each atom averaged for all tokens in SMILES and all heads was visualized. Note that we did not study about the decoder because some attention combinations were masked.
Attention to tokens subordinate to one atom (like '[', ']', and '@') was added to the attention to that atom,

while attention to tokens which cannot be attributed to a certain atom (like '-' or '=' representing bonds) was ignored.

## 4 Results and Discussion

### 4.1 Generative and reconstruction performance of Transformer VAE

The Transformer VAE model was trained with the MOSES train dataset, and 30,000 SMILES were generated by the trained decoder from random latent variables following the prior distribution. **Table 1** shows the property of the generated molecules averaged for 3 trials compared to those from other generative models trained with the MOSES dataset. The results showed that while the Transformer VAE model maintains high validity and uniqueness, it showed a high ability to generate molecules which do not exist in the training set. The high novelty and uniqueness imply that the current model succeeded in encompassing far more molecules in the latent space, suggesting that the chemical space generated by the model is useful for new drug discovery. The model also showed high scaffold similarity with the testSF set, which have scaffolds not seen in the train set. This means that our model has greater ability to generate novel scaffolds, which can be the reason for the high novelty of the generated molecules. This feature of the proposed model suggests the possibility of further study to screen novel drug candidates in the latent space using docking simulator or its emulator [41].

Reconstruction performance of the molecules in the test set were also measured, and *perfect accuracy* (ratio of completely matched predictions) and *partial accuracy* (character-wise correspondence of the predicted strings, see methods for both metrics) reached nearly 90% and 99%, respectively (**Table 2**). This indicates that about 90% of molecules were reversibly mapped to the latent space, and it is suggested that the model can also be used for generation of molecular descriptor, and that the tasks related to the distribution of molecules can be performed in the latent space before decoding, such as molecule optimization, active learning and definition of applicability domain of machine learning models.

Distribution of the predicted mean and variance of the latent variables for all elements of latent variables and all molecules in test set and all molecules in the test were shown in **Figure 2**, and it shows that most of the elements in the latent variables follow normal distribution $\mathcal{N}(0, I)$ (same as prior distribution), while a few elements have low variance. This suggests that the structural information of the input is held by only a few elements, while most of the elements have no information. We conducted further experiment to confirm this in section 4.2.4

### 4.2 Ablation studies

4.2.1 Training length

We investigated the learning progress and generative performance of the Transformer VAE model during training for 3 trials. One of the trials was prolonged to 1M steps to observe the saturation of the performance. The results indicated that, over the course of training, the Transformer VAE model gradually learned to generate molecules similar to the training set, while the ability to generate novel scaffolds (measured by scaffold similarity to the testSF set) slightly decreased (**Figure 3a**). The validity, a metric indicating the property of generating valid chemical structures, consistently remained high during the learning progress. These findings suggest that adjusting the length of training is crucial, depending on whether the goal is to generate novel or similar molecules to the learned ones. In terms of reconstruction performance, partial accuracy saturated at early steps, while the convergence of perfect accuracy required longer training, consistent with a previous study (**Figure 3b**)[42]. The transition of the distribution of the latent space showed gradual convergence to the prior probability distribution ($\mu = 0, \sigma^2 = 1$), indicating that the model successfully balanced the reconstruction of inputs and regularization of the latent space during training (**Figure 3c and 3d**).

4.2.2 Weight of regularization on latent variables

The proposed Transformer VAE model set the ratio of $D_{KL}$ loss to reconstruction loss (defined as $\beta$) to 0.01. Here this ratio was varied from 0 (no regularization) to 1 (normal VAE) to investigate the effect of $\beta$. When $\beta$ is 0, the latent variables are not regularized. In this case, the latent variables to be decoded into SMILES were generated from a normal distribution, where the distribution has the same mean and variance as the latent variables of the sampled training set. The results showed that the validity and scaffold similarity to the test set of the generated molecules improved as $\beta$ increased (except when $\beta$ is 1), while novelty and scaffold similarity to the testSF set deteriorated (**Figure 4a**). These findings suggest that a larger $\beta$ encourages the model to generate molecules close to the learned chemical space, while smaller $\beta$ values result in the generation of more novel molecules. Large $\beta$ can degrade the reconstruction performance of the model, but regularization of latent variables has little effect on reconstruction when $\beta$ is small (~0.001), compared with the performance of the model without regularization ($\beta$=0) (**Figure 4b**). The distribution of mean and variance of the latent variables indicated that the distribution becomes closer to the standard normal distribution as $\beta$ is increased, while a few elements continue to hold molecular information with low variance, except when $\beta$ is 1 (**Figure 4c and 4d**). Posterior collapse was observed when $\beta$ is 1 and the model returns to a normal VAE, resulting in the complete correspondence of the distribution of the latent space and prior distribution, and low novelty of generated molecules.

4.2.3 Comparison with other generative model architecture

To assess the impact of the introduction of VAE, a comparison was made with two different architectures based on Transformer (**Figure 5**). The first model, named *no-reparametrization*, simply decodes SMILES from latent variables generated by the encoder without reparametrization by the normal distribution. This model resembles the VAE model when $\beta = 0$. The second model is a generative language model, which learns the distribution of each token in SMILES conditioned by the preceding tokens. This model learns the distribution of molecules like a collapsed VAE model. For the language model, SMILES strings were generated not by beam search but by sampling each token following the predicted distribution. This generation method was also applied to the VAE model for comparison. The results indicated that the VAE model outperformed the other models in generating novel molecules with novel scaffolds (**Table 3**). This outcome suggests the advantage of VAE in generating novel data compared to a language model, which is currently prevalent in generative tasks.

4.2.4 Dimension of latent variables

**Figure 2** suggests that the structural information of molecules is primarily encoded by a few elements of latent variables with low variance, while most of the elements follow the prior distribution. To explore this further, the dimension of latent variables was varied, and the generative and reconstruction performance was compared. The results revealed that generative performance remained consistent until the dimension was decreased to 16, and all metrics, except scaffold similarity to the test set, were unaffected when $\beta$ was decreased to 8 (**Figure 6a**). Reconstruction accuracy also remained at a consistent level when the dimension was 16 or larger (**Figure 6b and 6c**). These findings indicate that the structural information of molecules and the diversity of the MOSES dataset can be adequately represented by 16-dimensional latent variables. This dimension is considerably shorter than commonly used molecular descriptors such as ECFP (~2048 dimensions) or MACCS (166 or more dimensions), as well as deep-learning-based descriptors like CDDD[26] (512 dimensions). This suggests the possibility of using more lightweight machine learning models in tasks related to molecules. The required dimension of latent variables was found to depend on the dataset to be learned, as discussed in the following section.

## 4.4 Training with ZINC-15 dataset

We have been training Transformer VAE with the MOSES dataset which satisfies rigid conditions regarding drug likeliness, but a model trained on a larger dataset could generate various and novel molecules that contribute to the discovery of new drug seeds. To examine the ability of the proposed model to generate novel molecules from a larger dataset, we trained the model using the ZINC-15 dataset, which contains approximately 1 billion molecules. The model was trained with the same hyperparameters as the training with the MOSES dataset and successfully converged without posterior collapse (**Supplementary Table 5**, **Figure 7a and 7b**). The generated molecules showed high uniqueness, validity and novelty (**Table 4**) when generating SMILES.

When the dimension of the latent variables was varied, the reconstruction performance did not change when the dimension was 32 or larger (**Figure 7c**). However, unlike the model trained by the MOSES dataset, perfect accuracy decreased when the dimension was 16. This result suggests that the molecules in the ZINC-15 dataset are more diverse and require a higher dimension (~32) for adequate description, although the required dimension is still smaller than that of commonly used descriptors. This also suggests that the diversity of dataset needs to be considered to decide proper model size.

### 4.5 Molecular property prediction from latent variables

The latent variables encoded by the model are expected to contain structural information of molecules, and can be used as molecular descriptors like those of other chemical language models [6,26]. We therefore attempted to predict the molecular property and activity data provided by MoleculeNet from the mean of the latent variables estimated by the encoder (before reparameterization). **Figure 8** shows the prediction performance compared with existing molecular descriptors and an end-to-end model (See methods for the details about them). The latent variables as a descriptor showed competitive performance against existing methods in predicting most of the molecular properties. These results indicate the capacity of the model to learn joint distribution of the output and molecular properties, and to generate molecules conditioned by molecular properties. Notably, each variable is derived from normal distribution, which would provide us the advantage of facilitating the consideration and definition of applicability domain when latent representations are used as descriptors for molecular property prediction.

### 4.6 Visualization of attention

Finally, attention weights to each token were visualized to clarify important substructures that were strongly attended by the other tokens. **Figure 9** shows the weight of the attention targeted to each token, averaged for all source tokens of attention and all heads. Note that only attention in the 1st, 4th, and 8th (last) layers is shown due to space restrictions. The results showed that rare atoms or substructures received relatively large amount of attention, and suggested that such information is emphasized in the latent variables. Thus, one of the advantages of this Transformer-based model is that important atoms can be visualized when generating latent variables, and this method can contribute to improving the accountability of chemical language models, particularly in settings where the models learn various chemical structures in pretraining (not end-to-end purpose).

## 5 Conclusion

The contributions of this study are as follows:

1. We have developed a Transformer-VAE model that learns diverse chemical structures without encountering posterior collapse.
2. The proposed model has demonstrated state-of-the-art performance in generating unseen molecules.
3. Ablation study implied the advantage of VAE in generating novel molecules, compared to other architectures such as a language model.
4. We showed that the molecules in MOSES or ZINC-15 dataset can be described by variables of only 16 to 32 elements and such dataset dependency indicated the importance of dataset-specific considerations in model design.
5. The latent representation of the VAE showed competitive performance in predicting molecular properties, suggesting the capacity for fast screening by the association of molecular properties with latent representation.
6. Attention weight in VAE encoder was visualized to show important structure in generating latent representation of molecules by Transformer VAE model.

The novelty of this model is that it combines Transformer with VAE in the field of molecule generation. The model structure and hyperparameters were optimized to improve the performance of molecule generation. The generated molecules can be searched for various novel drug seeds by virtual screening. Recently, there has been a growing demand for Virtual Chemical Libraries (VCL), particularly in light of advancements in virtual screening methodologies. Our proposed model offers the capability to easily generate a customized VCL from Gaussian random numbers and has the potential to access unseen chemical spaces.

A major limitation of our model is the lack of consideration of synthesis difficulty and *in vivo* stability. This limitation can potentially be addressed by integrating models that evaluate synthesis complexity or implementing filtering strategies as a virtual screening pipeline[43,44]. Thus, we believe that our model contributes to the foundational aspects of drug discovery through the use of deep generative models, particularly in handling diverse chemical structures.

## Author Contribution

Yasuhiro Yoshikai: Methodology, Software, Investigation, Writing – Original Draft, Visualization.
Tadahaya Mizuno: Conceptualization, Resources, Supervision, Project administration, Writing – Original Draft, Writing – Review & Editing, Funding acquisition.
Shumpei Nemoto: Methodology, Software
Hiroyuki Kusuhara: Writing – Review

## Conflicts of Interest

The authors declare that they have no conflicts of interest.

## Availability

Code and models are available at https://github.com/mizuno-group/TransformerVAE.

## Acknowledgement

We thank all those who contributed to the construction of the following data sets employed in the present study such as ZINC-15 and MoleculeNet. This work was supported by AMED under Grant Number JP22mk0101250h and 23ak0101199h0001. This work used computational resources of supercomputer Wisteria provided by Information Technology Center (The University of Tokyo).

## References

1. Polishchuk, P. G., Madzhidov, T. I. & Varnek, A. Estimation of the size of drug-like chemical space based on GDB-17 data. *J Comput Aided Mol Des* **27**, 675–679 (2013).
2. Sterling, T. & Irwin, J. J. ZINC 15–ligand discovery for everyone. *J Chem Inf Model* **55**, 2324–2337 (2015).
3. Kim, S. *et al.* PubChem substance and compound databases. *Nucleic Acids Res* **44**, D1202–D1213 (2016).
4. Mendez, D. *et al.* ChEMBL: towards direct deposition of bioassay data. *Nucleic Acids Res* **47**, D930–D940 (2019).
5. Kingma, D. P. & Welling, M. Auto-encoding variational bayes. Preprint at https://arxiv.org/abs/1312.6114 (2013).
6. Gómez-Bombarelli, R. *et al.* Automatic Chemical Design Using a Data-Driven Continuous Representation of Molecules. *ACS Cent Sci* **4**, 268–276 (2018).
7. Kusner, M. J., Paige, B. & Miguel Hernández-Lobato, J. *Grammar Variational Autoencoder*. http://opensmiles.org/spec/open-smiles-2-grammar.html (2017).
8. Vaswani, A. *et al.* Attention Is All You Need. in *Advances in Neural Information Processing Systems* (2017).
9. Devlin, J., Chang, M.-W., Lee, K., Kristina, T. & Language, A. I. BERT: Pre-training of Deep Bidirectional Transformers for Language. Preprint at https://arxiv.org/abs/1810.04805 (2018).
10. Openai, A. R., Openai, K. N., Openai, T. S. & Openai, I. S. *Improving Language Understanding by Generative Pre-Training*. https://gluebenchmark.com/leaderboard.
11. Dollar, O., Joshi, N., Beck, D. A. C. & Pfaendtner, J. Attention-based generative models for: De novo molecular design. *Chem Sci* **12**, 8362–8372 (2021).
12. Bowman, S. R. *et al.* Generating Sentences from a Continuous Space. (2015).
13. Fang, L. *et al.* Transformer-based Conditional Variational Autoencoder for Controllable Story Generation. (2021).
14. Alemi, A. A. *et al.* Fixing a Broken ELBO. (2017).
15. Jin, W., Barzilay, R. & Jaakkola, T. *Junction Tree Variational Autoencoder for Molecular Graph Generation*. (2018).
16. Liu, Q., Allamanis, M., Brockschmidt, M. & Gaunt, A. L. *Constrained Graph Variational Autoencoders for Molecule Design*.

17. *Deep Learning for Molecular Design-a Review of the State of the Art.*
18. Lim, J., Ryu, S., Kim, J. W. & Kim, W. Y. Molecular generative model based on conditional variational autoencoder for de novo molecular design. *J Cheminform* **10**, (2018).
19. Chenthamarakshan, V. *et al. CogMol: Target-Specific and Selective Drug Design for COVID-19 Using Deep Generative Models.*
20. Samanta, S., O'Hagan, S., Swainston, N., Roberts, T. J. & Kell, D. B. VAE-Sim: A novel molecular similarity measure based on a variational autoencoder. *Molecules* **25**, (2020).
21. Bagal, V., Aggarwal, R., Vinod, P. K. & Priyakumar, U. D. MolGPT: molecular generation using a transformer-decoder model. *J Chem Inf Model* **62**, 2064–2076 (2021).
22. Hong, Y.-B., Lee, K.-J., Heo, D. & Choi, H. Molecule Generation for Drug Discovery with New Transformer Architecture. Preprint at https://ssrn.com/abstract=4195528 (2022).
23. Wang, J. *et al.* Multi-constraint molecular generation based on conditional transformer, knowledge distillation and reinforcement learning. *Nat Mach Intell* **3**, 914–922 (2021).
24. Cofala, T. & Kramer, O. *Transformers for Molecular Graph Generation.*
25. Kim, H., Na, J. & Lee, W. B. Generative chemical transformer: neural machine learning of molecular geometric structures from chemical language via attention. *J Chem Inf Model* **61**, 5804–5814 (2021).
26. Winter, R., Montanari, F., Noé, F. & Clevert, D.-A. Learning continuous and data-driven molecular descriptors by translating equivalent chemical representations. *Chem Sci* **10**, 1692–1701 (2019).
27. Xiong, R. *et al. On Layer Normalization in the Transformer Architecture.* (2020).
28. Williams, R. J. & Zipser, D. *A Learning Algorithm for Continually Running Fully Recurrent Neural Networks.* http://direct.mit.edu/neco/article-pdf/1/2/270/811849/neco.1989.1.2.270.pdf.
29. Polykovskiy, D. *et al.* Molecular Sets (MOSES): A Benchmarking Platform for Molecular Generation Models. *Front Pharmacol* **11**, (2020).
30. Irwin, R., Dimitriadis, S., He, J. & Bjerrum, E. J. Chemformer: A pre-trained transformer for computational chemistry. *Mach Learn Sci Technol* **3**, 015022 (2022).
31. Bjerrum, E. J. & Sattarov, B. Improving chemical autoencoder latent space and molecular de novo generation diversity with heteroencoders. *Biomolecules* **8**, (2018).
32. Yoshikai, Y., Mizuno, T., Nemoto, S. & Kusuhara, H. *Difficulty in Learning Chirality for Transformer Fed with SMILES.*
33. RDKit.
34. Nemoto, S., Mizuno, T. & Kusuhara, H. Investigation of chemical structure recognition by encoder–decoder models in learning progress. *J Cheminform* **15**, (2023).
35. Wu, Z. *et al.* MoleculeNet: A benchmark for molecular machine learning. *Chem Sci* **9**, 513–530 (2018).
36. Ramsundar, B. *MOLECULAR MACHINE LEARNING WITH DEEPCHEM.* http://purl.stanford.edu/js264hd4826 (2018).
37. Chen, T. & Guestrin, C. XGBoost: A Scalable Tree Boosting System. (2016) doi:10.1145/2939672.2939785.
38. Akiba, T., Sano, S., Yanase, T., Ohta, T. & Koyama, M. Optuna: A next-generation hyperparameter optimization framework. in *Proceedings of the 25th ACM SIGKDD international conference on knowledge discovery & data mining* 2623–2631 (2019).
39. Rogers, D. & Hahn, M. Extended-connectivity fingerprints. *J Chem Inf Model* **50**, 742–754 (2010).
40. Zhou, G. *et al.* Uni-Mol: a universal 3D molecular representation learning framework. in *International Conference on Learning Representations* (2023).
41. Gentile, F. *et al.* Deep Docking: A Deep Learning Platform for Augmentation of Structure Based Drug Discovery. *ACS Cent Sci* **6**, 939–949 (2020).
42. Yoshikai, Y., Mizuno, T., Nemoto, S. & Kusuhara, H. Difficulty in chirality recognition for Transformer architectures learning chemical structures from string representations. *Nat Commun* **15**, 1197 (2024).
43. Brenk, R. *et al.* Lessons learnt from assembling screening libraries for drug discovery for neglected diseases. *ChemMedChem* **3**, 435–444 (2008).

44. Baell, J. B. & Holloway, G. A. New substructure filters for removal of pan assay interference compounds (PAINS) from screening libraries and for their exclusion in bioassays. *J Med Chem* **53**, 2719–2740 (2010).

## Figures and Tables

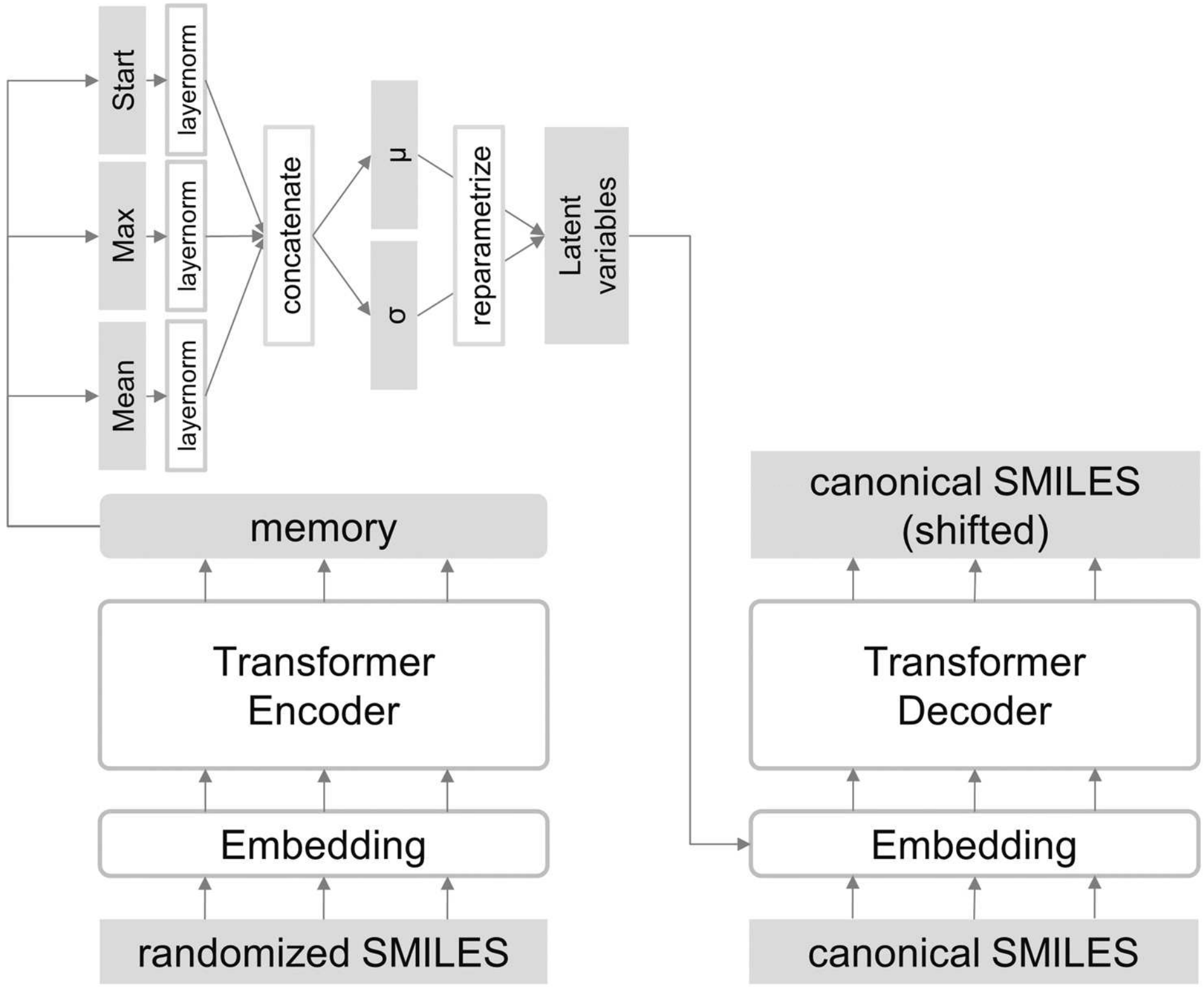

**Figure 1. Structure of Transformer VAE model**

Visual summary of the model proposed in this paper. Filled shapes represent variables, and white shapes represent layers of process. Randomized SMILES was inputted into the encoder, and distribution of latent variables $z$ were estimated from the pooled memory. Latent variables were added to the embedding of canonical SMILES to be decoded. Teacher forcing was used in the decoder during training, while beam search was used to generate new molecules.

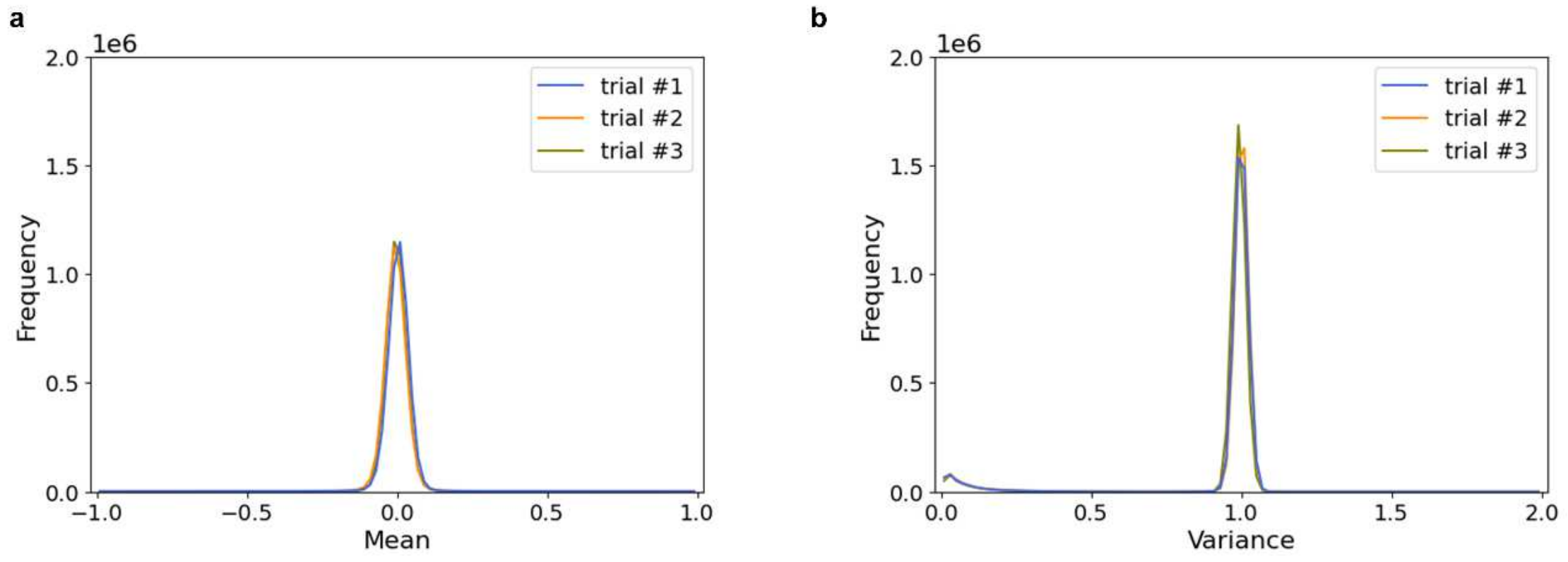

**Figure 2. Distribution of the predicted mean and variance of latent variables**

(a, b) Distribution of mean and variance of latent variables estimated by the VAE encoder, for all elements and all molecules in test set at the end of the training for 3 training trials. Bin length was 0.02 for both mean and variance, and frequency indicates the number of elements in each bin.

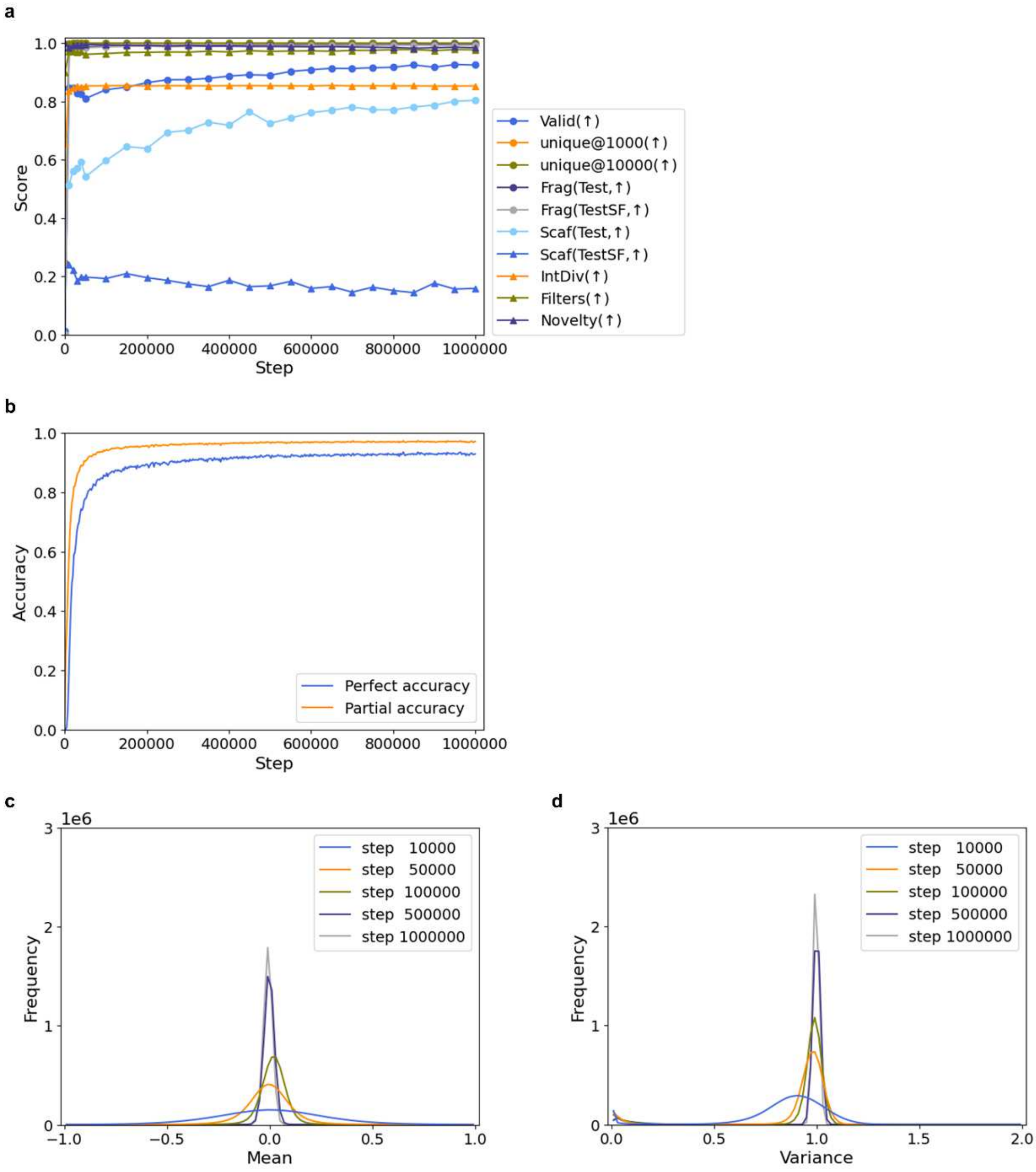

**Figure 3. Effect of training steps**

(a) Temporal change of generative performance of the model evaluated by the metrics provided by the dataset. See Method section for detailed explanation of each metrics. The training was conducted for 1000000 steps, and the performance was evaluated at various training steps. (b) Temporal change of perfect/partial accuracy for 1000000 steps. Perfect accuracy is the ratio of SMILES in training set which the model was able to decode completely, and partial accuracy is the average ratio of the decoded tokens which matched the original. Greedy decode was used to generate prediction. (c, d) Distribution of mean and variance of latent variables estimated by the VAE encoder, for all elements and all molecules in test set at different training steps. Bin length was 0.02 for both mean and variance, and frequency indicates the number of elements in each bin.

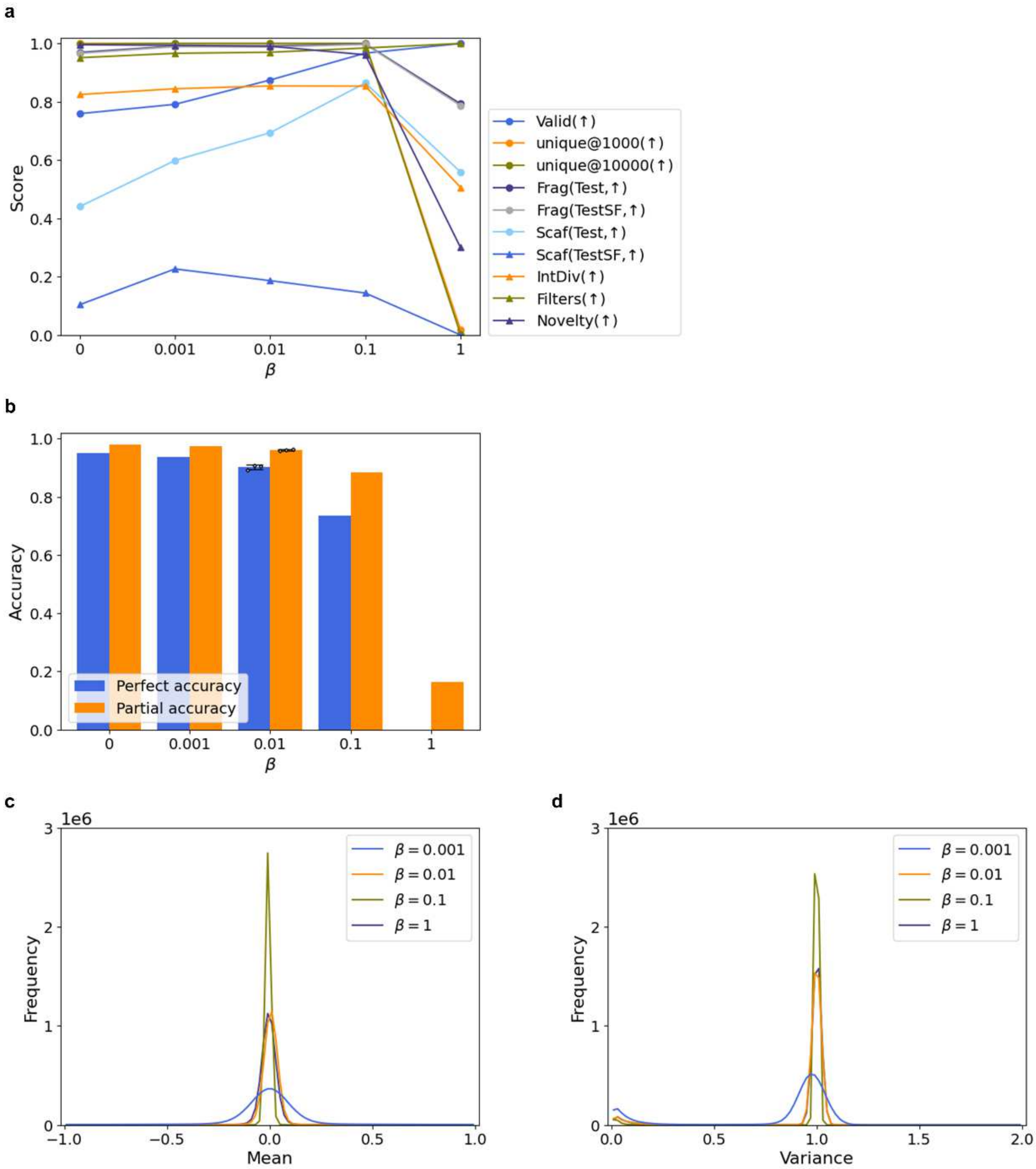

**Figure 4. Effect of weight on $D_{KL}$**

(a) Generative performance of the model evaluated by the metrics provided by the dataset at the end of the training with different weight of loss on $D_{KL}$ to reconstruction loss (defined as $\beta$). See Method section for detailed explanation of each metrics. (b) Perfect accuracy / partial accuracy of the model trained with different $\beta$. Perfect accuracy is the ratio of SMILES in training set which the model was able to decode completely, and Partial accuracy is the average ratio of the decoded tokens which matched the original. Greedy decode was used to generate prediction. (c, d) Distribution of mean and variance of latent variables estimated by the VAE encoder, for all elements and all molecules in test set at the end of the training with different $\beta$. Bin length was 0.02 for both mean and variance, and frequency indicates the number of elements in each bin.

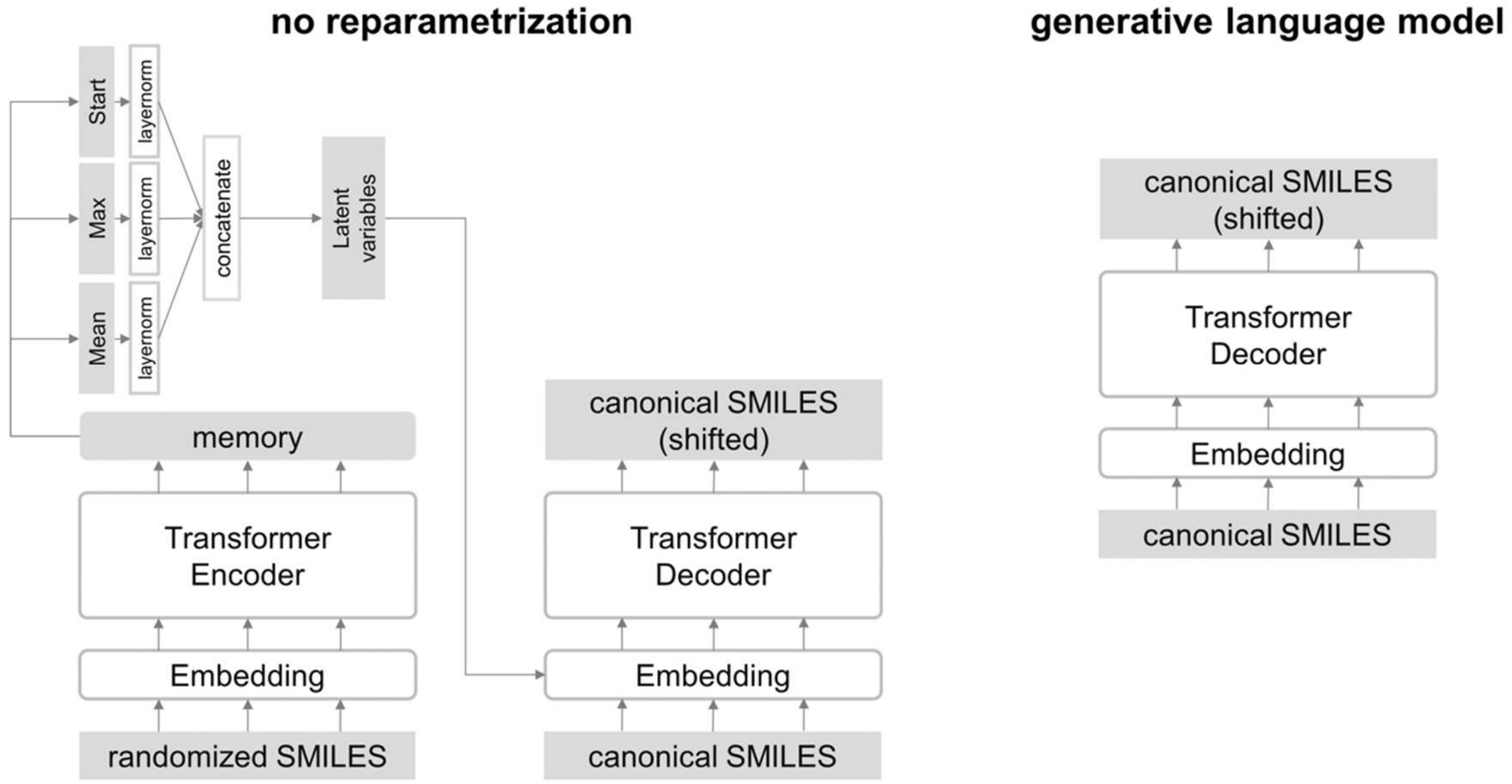

**Figure 5. The structure of the model compared with VAE**

Structure of two Transformer-based models used as baselines of generative performance. *No-reparametrization* model has similar structure as VAE, but it decodes SMILES from the latent variables outputted from the encoder without reparametrization, and generates molecules from randomly generated latent variables following normal distribution with the same mean and variance of randomly sampled training set. *Language model* estimates the distribution of each token conditioned by the preceding tokens, and it generates SMILES by sampling tokens one by one following the estimated distribution.

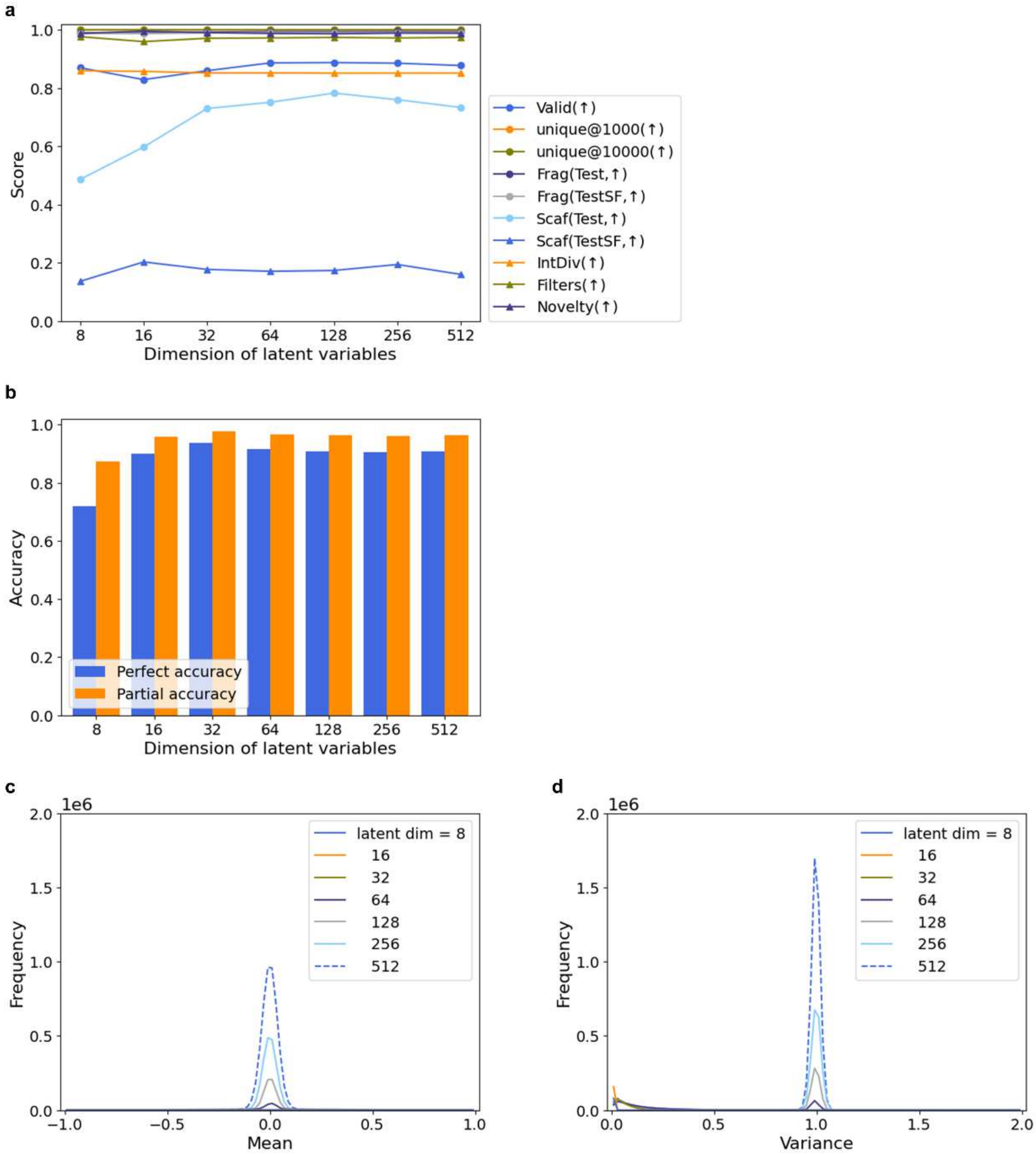

**Figure 6. Effect of the dimension of the latent variables**

(a) Generative performance of the model evaluated by the metrics provided by the dataset at the end of the training with different dimension of latent variables. See Method section for detailed explanation of each metrics. (b) Perfect accuracy / partial accuracy with different dimension of latent variables. Perfect accuracy is the ratio of SMILES in training set which the model was able to decode completely, and Partial accuracy is the average ratio of the decoded tokens which matched the original. Greedy decode was used to generate prediction. (c, d) Distribution of mean and variance of latent variables estimated by the VAE encoder, for all elements and all molecules in test set at the end of the training with different dimension of latent variables. Bin length was 0.02 for both mean and variance, and frequency indicates the number of elements in each bin.

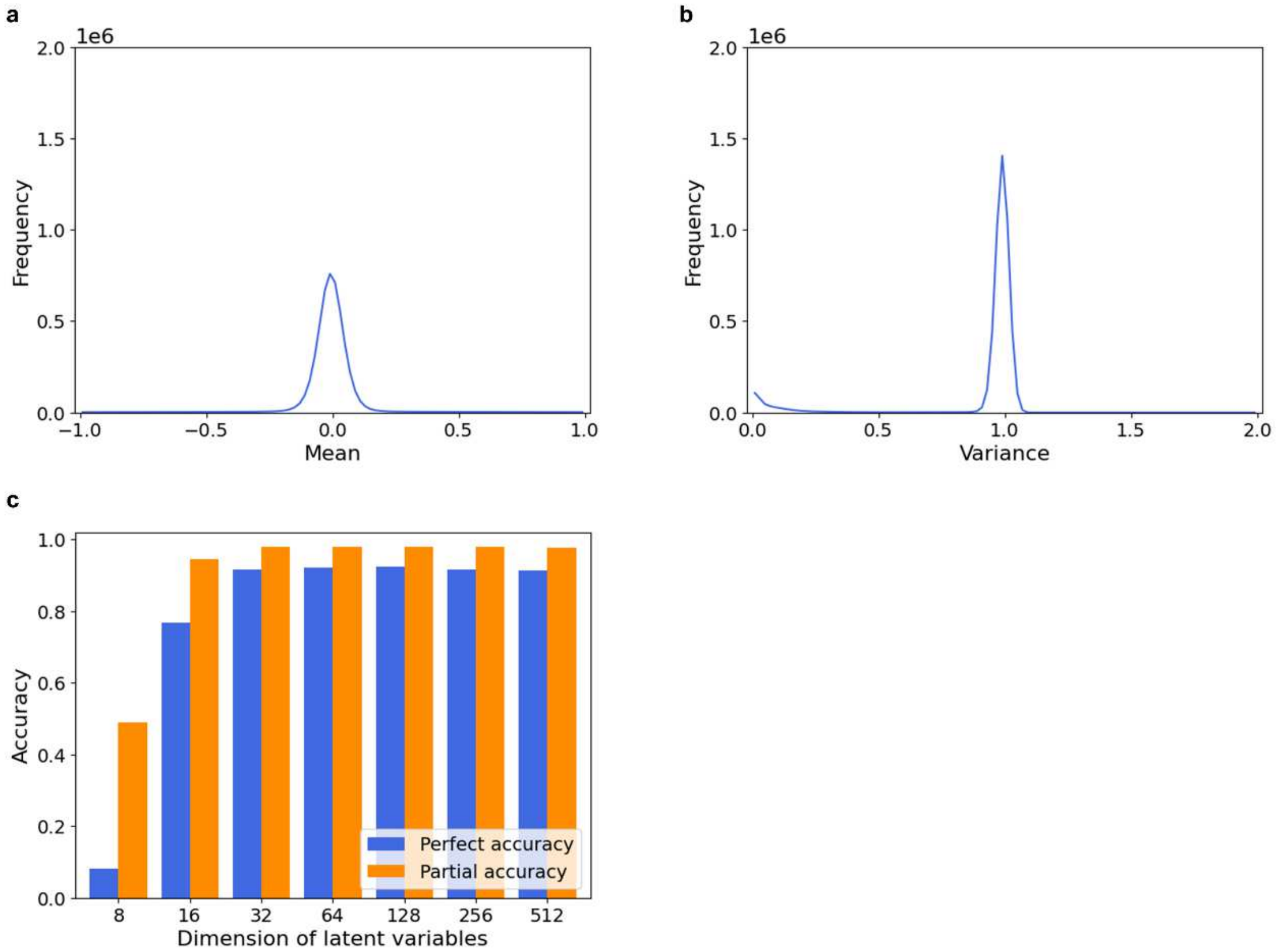

**Figure 7. Distribution of latent variables when the model was trained with ZINC-15 dataset**

(a, b) Distribution of mean and variance of latent variables estimated by the VAE encoder, for all dimensions and all molecules in test set at the end of the training with ZINC-15 dataset. Bin length was 0.02 for both mean and variance, and frequency indicates the number of elements in each bin. (c)

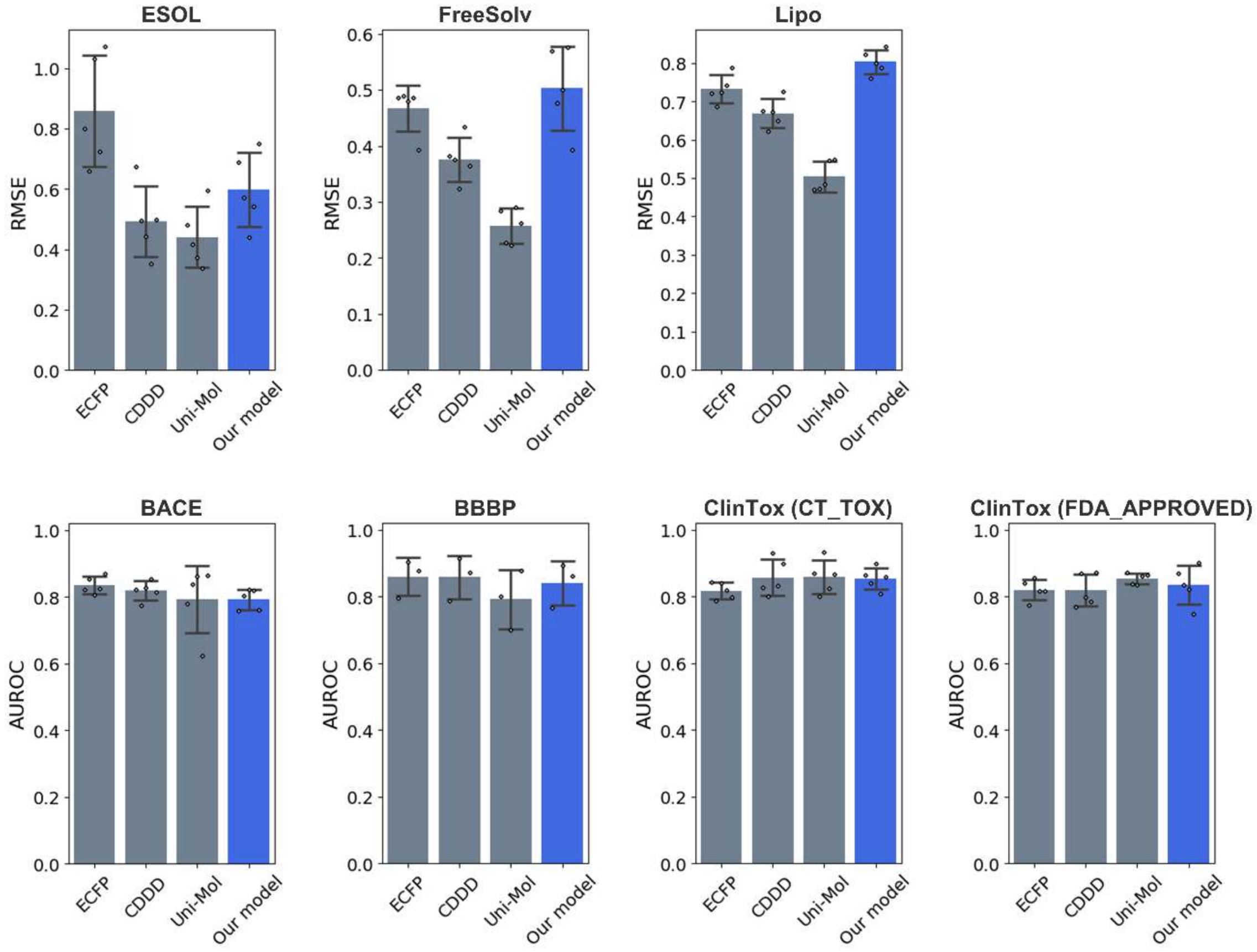

**Figure 8. Performance of molecular property prediction by latent variables**

Performance of molecular property prediction by mean of latent variables estimated by the encoder from the input randomized SMILES. The metrics of the prediction performance were decided based on MoleculeNet. ECFP and CDDD was used as baseline descriptor of molecules, and Uni-Mol was used as baseline model. For all descriptors, XGBoost was used as predictor and hyperparameters were optimized by Bayesian optimization with optuna. Prediction was conducted and evaluated for 5 folds splitted by the methods recommended by MoleculeNet.

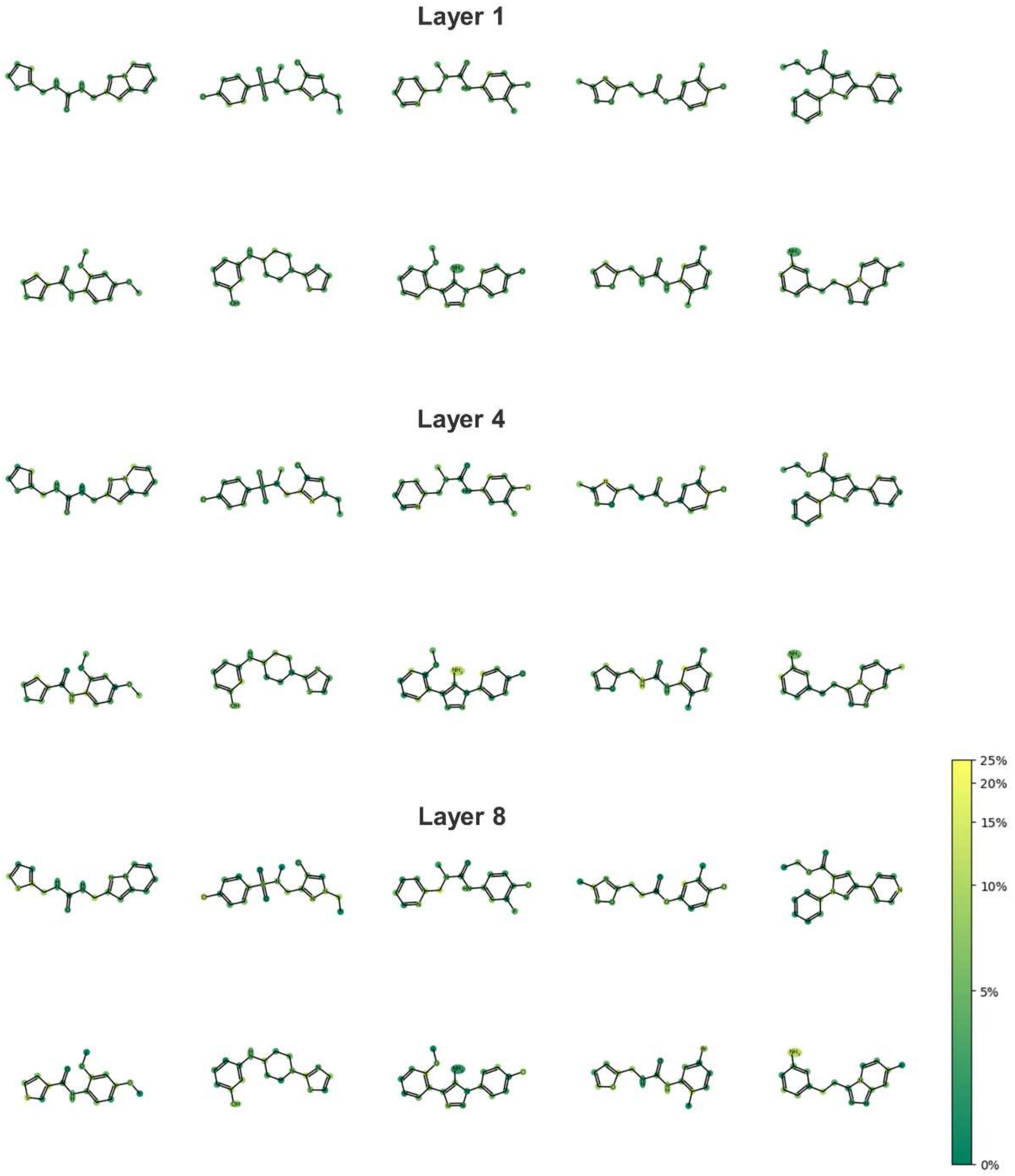

**figure 9. Visualized weight of attention on each atom.**

Weights of attention to each atom averaged for all sources of attention in several layers of the encoder. Attention to tokens which are not element symbols but belong to certain atoms is incorporated into those atoms, while attention to tokens representing bonds or other grammatical information are ignored. Nonlinear colormap is used to highlight the difference of weights.

**Table 1. Generative performance of the Transformer VAE model, compared to baseline models in MolecularSets**

(↑)/ (↓) means higher/lower value is better.

| | | **Our model** | HMM | NGram | Combinatorial | CharRNN | AAE | VAE | JTN-VAE | LatentGAN |
|---|---|---|---|---|---|---|---|---|---|---|
| Valid(↑) | | 0.8761±0.0035 | 0.0760±0.0322 | 0.2376±0.0025 | **1.0000±0.0000** | 0.9748±0.0264 | 0.9368±0.0341 | 0.9767±0.0012 | **1.0000±0.0000** | **0.8966±0.0029** |
| unique@1000(↑) | | **1.0000±0.0000** | 0.6230±0.1224 | 0.9740±0.0108 | 0.9983±0.0015 | **1.0000±0.0000** | **1.0000±0.0000** | **1.0000±0.0000** | **1.0000±0.0000** | **1.0000±0.0000** |
| unique@10000(↑) | | **1.0000±0.0001** | 0.5671±0.1424 | 0.9217±0.0019 | 0.9909±0.0009 | 0.9994±0.0003 | 0.9973±0.0020 | 0.9984±0.0005 | 0.9996±0.0003 | |
| FCD(↓) | Test | 1.3380±0.0735 | 24.4661±2.5251 | 5.5069±0.1027 | 4.2375±0.0370 | **0.0732±0.0247** | 0.5555±0.2033 | 0.0990±0.0125 | | 0.2968±0.0087 |
| | TestSF | 1.7098±0.0740 | 25.4312±2.5599 | 6.2306±0.0966 | 4.5113±0.0274 | **0.5204±0.0379** | 1.0572±0.2375 | 0.5670±0.0338 | | 0.8281±0.0117 |
| SNN(↑) | Test | 0.4803±0.0025 | 0.3876±0.0107 | 0.5209±0.0010 | 0.4514±0.0003 | 0.6015±0.0206 | 0.6081±0.0043 | **0.6257±0.0005** | 0.5477±0.0076 | 0.5371±0.0004 |
| | TestSF | 0.4658±0.0026 | 0.3795±0.0107 | 0.4997±0.0005 | 0.4388±0.0002 | 0.5649±0.0142 | 0.5677±0.0045 | **0.5783±0.0008** | 0.5194±0.0070 | 0.5132±0.0002 |
| Frag(↑) | Test | 0.9927±0.0016 | 0.5754±0.1224 | 0.9846±0.0012 | 0.9912±0.0004 | **0.9998±0.0002** | 0.9910±0.0051 | 0.9994±0.0001 | | 0.9986±0.0004 |
| | TestSF | 0.9898±0.0018 | 0.5681±0.1218 | 0.9815±0.0012 | 0.9904±0.0003 | 0.9983±0.0003 | | **0.9985±0.0003** | 0.9947±0.0002 | 0.9972±0.0007 |
| Scaf(↑) | Test | 0.7159±0.0192 | 0.2065±0.0481 | 0.5302±0.0163 | 0.4445±0.0056 | 0.9242±0.0058 | | **0.9386±0.0021** | 0.8964±0.0039 | 0.8867±0.0009 |
| | TestSF | **0.1893±0.0081** | 0.0490±0.0180 | 0.0977±0.0142 | 0.0865±0.0027 | 0.1101±0.0081 | | 0.0588±0.0095 | 0.1009±0.0105 | 0.1072±0.0098 |
| IntDiv(↑) | | 0.8531±0.0013 | 0.8466±0.0403 | **0.8738±0.0002** | 0.8732±0.0002 | | 0.8557±0.0031 | 0.8558±0.0004 | 0.8551±0.0034 | 0.8565±0.0007 |
| Filters(↑) | | 0.9706±0.0005 | 0.9024±0.0489 | 0.9582±0.0010 | 0.9557±0.0018 | 0.9943±0.0034 | 0.9960±0.0006 | **0.9970±0.0002** | 0.9760±0.0016 | 0.9735±0.0006 |
| Novelty(↑) | | 0.9911±0.0005 | **0.9994±0.0010** | 0.9694±0.0010 | 0.9878±0.0008 | 0.8419±0.0509 | 0.7931±0.0285 | 0.6949±0.0069 | 0.9143±0.0058 | 0.9498±0.0006 |

**Table 2. Reconstruction performance of the model in 3 training trials**

| | Perfect accuracy | Partial accuracy |
|---|---|---|
| trial #1 | 0.8927 | 0.9571 |
| trial #2 | 0.9067 | 0.9621 |
| trial #3 | 0.9065 | 0.9632 |
| mean±std | 0.9020±0.0080 | 0.9608±0.0032 |

**Table 3. Generative performance of the Transformer VAE model, compared to other architectures**

| | | VAE (beam search) | VAE (token sampling) | no-reparametrization | language model |
|---|---|---|---|---|---|
| Valid( ↑ ) | | 0.8747 | | 0.7735 | **0.9960** |
| unique@1000(↑) | | **1.0000** | **1.0000** | 0.9990 | **1.0000** |
| unique@10000(↑) | | **1.0000** | **1.0000** | 0.9977 | 0.9980 |
| FCD( ↓ ) | Test | 1.4128 | | 6.1512 | **0.0712** |
| | TestSF | 1.7952 | 1.8560 | 6.2894 | **0.5459** |
| SNN( ↑ ) | Test | 0.4774 | **0.4741** | 0.4936 | 0.6417 |
| | TestSF | 0.4628 | **0.4603** | 0.4811 | 0.5871 |
| Frag( ↑ ) | Test | 0.9911 | 0.9903 | 0.9611 | **0.9998** |
| | TestSF | 0.9880 | | 0.9577 | **0.9982** |
| Scaf(↑) | Test | 0.6939 | | 0.3726 | **0.9394** |
| | TestSF | 0.1863 | **0.1914** | 0.1235 | 0.0321 |
| IntDiv( ↑ ) | | 0.8544 | 0.8547 | | **0.8571** |
| Filters( ↑ ) | | 0.9700 | | 0.9618 | **0.9990** |
| Novelty( ↑ ) | | 0.9906 | **0.9916** | 0.9942 | 0.4800 |

**Table 4. Scores of molecules generated by Transformer VAE model trained with ZINC-15 dataset.**

(↑)/ (↓) means higher/lower value is better.

| Metric | Score |
|---|---|
| Valid(↑) | 0.9010 |
| unique@1000(↑) | 1.0000 |
| unique@10000(↑) | 1.0000 |
| IntDiv(↑) | 0.8518 |
| Novelty(↑) | 0.9675 |

# Supplementary Information for
# "A novel molecule generative model of VAE combined with Transformer"

## Supplementary Tables

**Supplementary Table 1. Hyperparameters of the Transformer VAE model and training**

| Parameter | Value |
|---|---|
| n_layer | 8 |
| d_model | 512 |
| dim_feedforward | 2048 |
| dropout | 0 |
| scheduler | warmup scheduler[8] |
| warmup step | 4000 |
| max learning rate | 0.001 |

**Supplementary Table 2. Tokens in Transformer VAE model**

| | |
|---|---|
| Special tokens | <s>, </s>, <pad> |
| Normal tokens | 0 1 2 3 4 5 6 7 8 9 ( ) [ ] : = @ @@ + / . - # % b c n o s p H B C N O S P F Cl Br I |

**Supplementary Table 3. Datasets used for molecular property prediction**

Number of valid molecules is the number of SMILES which successfully generated molecules, and organic molecules are those which contains only organic atoms.

| Dataset | Task | Splitting | Metric | # of molecules | | |
|---|---|---|---|---|---|---|
| | | | | Total | Valid | Organic |
| ESOL | Regression | scaffold | RMSE | 1128 | 1128 | 1128 |
| FreeSolv | Regression | random | RMSE | 642 | 642 | 642 |
| Lipo | Regression | scaffold | RMSE | 4200 | 4200 | 4198 |
| BACE | Classification | scaffold | AUROC | 1513 | 1513 | 1513 |
| BBBP | Classification | scaffold | AUROC | 2050 | 2039 | 2039 |
| ClinTox | Classification | random | AUROC | 1484 | 1478 | 1456 |

**Supplementary Table 4. Search range of hyperparameters of XGBoost**

See documentation of XGBoost module (https://xgboost.readthedocs.io/en/stable) for explanation of each parameter.

| | Minimum | Maximum |
|---|---|---|
| eta | $1.0^{-8}$ | 10.0 |
| gamma | $1.0^{-8}$ | 10.0 |
| max_depth | 3 | 15 |
| min_child_weight | 1 | 10 |
| max_delta_step | 1 | 10 |
| subsample | 0.6 | 1.0 |
| colsample_bytree | 0.6 | 1.0 |
| lambda | $1.0^{-8}$ | 0.1 |
| alpha | $1.0^{-8}$ | 1.0 |

**Supplementary Table 5. Reconstruction performance of the model trained with ZINC-15 dataset**

| Perfect accuracy | Partial accuracy |
| --- | --- |
| 0.8775 | 0.9760 |